\documentclass[prb,english,twocolumn,amsmath,amssymb,superscriptaddress]{revtex4-2}
\usepackage{amsmath}
\usepackage{graphicx}
\usepackage{amssymb}
\usepackage{dsfont}
\usepackage{color}
\usepackage[bookmarks=true,colorlinks,linkcolor=blue,urlcolor=blue,citecolor=blue]{hyperref}

\newcommand{\new}[1]{\textcolor{black}{#1}}

\usepackage{hyperref}
\usepackage{tikz}
\usetikzlibrary{calc}

\newcommand{\be}{\begin{equation}}
\newcommand{\ee}{\end{equation}}
\newcommand{\bea}{\begin{eqnarray}}
\newcommand{\eea}{\end{eqnarray}}

\newcommand{\mc}{\mathcal}
\newcommand{\mb}{\mathbf}

\definecolor{darkorange}{rgb}{1, 0.55, 0.0}

\begin{document}

\title{Fractonic criticality in  Rydberg atom arrays}

\author{Rafael A. Mac\^{e}do}
\affiliation{Departamento de F\'{i}sica Te\'{o}rica e Experimental, Universidade Federal do Rio Grande do Norte, Natal, RN, 59078-970, Brazil}
\author{Rodrigo G. Pereira}
\affiliation{Departamento de F\'{i}sica Te\'{o}rica e Experimental, Universidade Federal do Rio Grande do Norte, Natal, RN, 59078-970, Brazil}
\affiliation{International Institute of Physics, Universidade Federal do Rio Grande do Norte, Natal, RN, 59078-970, Brazil}

\begin{abstract}
Fractonic matter can undergo unconventional  phase transitions driven by the condensation of   particles that move   along subdimensional manifolds. We propose that this type of  quantum critical point can be realized in a bilayer of crossed Rydberg chains. This system exhibits a transition between a disordered phase and a charge-density-wave phase with subextensive ground state degeneracy.  We show that this transition  is described by a stack of critical Ising conformal field theories which become \new{decoupled in the low-energy limit}. We also analyze the transition  using a Majorana mean-field approach for an effective lattice model, which confirms the picture of a fixed point of decoupled critical chains. We discuss the unusual scaling properties and derive anisotropic correlators that provide signatures of subdimensional criticality in this  realistic setup. 
\end{abstract}
\maketitle

\section{Introduction}
For decades, effective field theories have provided invaluable insight into quantum critical phenomena \cite{HertzPRB1976,SondhiRMP1997,sachdev2011quantum,fradkin2013field}. A general guiding principle is that   the low-energy properties of a   system close to a continuous phase transition  are governed by nearly massless excitations. If these excitations  propagate in all spatial directions,   one expects universal scaling behavior   once the correlation length diverges and   microscopic details  become irrelevant. 

The standard continuum limit    inherent in effective field theories   has recently been challenged by the   study of fracton phases  of matter  \cite{Nandkishore2019,pretko2020fracton,Gromov2024}.  Fractonic systems are characterized by  excitations that are completely immobile (fractons) or  propagate   along lower-dimensional subspaces (lineons and planons) \cite{ChamonPRL2005,BravyiAP2011,HaahPRA2011,VijayPRB2016}.  In addition, fracton phases  exhibit a subextensive ground state degeneracy.  These properties are associated with subsystem symmetries, which can be either exact or emergent \cite{LawlerPRB2004,nussinov2015compass,YouPRR2020,GorantlaPRB2021} and play an important role in quantum phase transitions \cite{MuhlhauserPRB2020,PoonPRR2021,ZhuPRL2023,LakePRB2021,RayhaunSciPost2023}.  In particular, continuous   transitions can be driven by the condensation of lineons and planons, leading to the notion of subdimensional criticality \cite{LakePRB2021,RayhaunSciPost2023}. In this intriguing scenario,  the transition is described by stacks of lower-dimensional critical theories,  \new{which   decouple  at low energies}.

In continuum descriptions of   fractons  \cite{SlaglePRB2017,PretkoPRB2018,BulmashPRB2018,MaPRB2018,GromovPRX2019,BurnellPRR2020,SeibergSciPost2020,SlaglePRL2021,FontanaSciPost2021,SullivanPRR2021},    physical observables often depend on a  short length scale related to a  lattice regularization. The need for this  regularization becomes apparent in theories with higher spatial derivatives and anisotropic scaling, an early example  of which  was  the Bose metal    in 2+1 dimensions  \cite{paramekanti2002exchange}; see also Ref. \cite{Seiberg2021}. In this case, the dispersion $\omega\sim k_xk_y$ of   bosonic spin modes vanishes along lines in momentum space. As a consequence,  high-momentum modes    contribute to the low-energy physics, a phenomenon known as UV-IR mixing \cite{SeibergSciPost2020,GorantlaPRB2021}. Similar theories have been proposed for  the fracton  critical point in a higher-order topological transition \cite{you2022fracton}, the fractonic Berezinskii-Kosterlitz-Thouless transition  in the   plaquette-dimer model \cite{YouMoessnerPRB2022,GrosvenorPRB2023}, and boundaries of   fracton models \cite{LuoPRB2022,fontana2023boundary}. Crucially,   UV-IR mixing imposes a modified  renormalization group (RG) analysis  \cite{paramekanti2002exchange,LawlerPRB2004,Xu2005,kapustin2018wilsonian,YouMoessnerPRB2022,LakePRB2022,GrosvenorPRB2023}. In some contexts,  the difficulties  with scaling  have been interpreted in terms of a dimensional reduction \cite{LawlerPRB2004,Xu2005}, whereby  the two-dimensional (2D) model is viewed as an array of 1D systems. Despite the enormous interest sparked by fracton-like physics, a major obstacle to  its observation  is that the proposed lattice models typically contain multi-spin interactions that are hard to realize experimentally.

 In this work, we show that a fractonic transition can be observed in a realistic setup with Rydberg atom arrays,  a   versatile platform for the quantum simulation of long-sought phases of matter \cite{BrowaeysNP2020,SemeghiniScience2021,SchollNature2021,SuracePRX2020,VerresenPRX2021,SlaglePRB2022,MyersonJainPRL2022,SamajdarPRL2023,LeePRL2023}.  Our setup consists of two layers of parallel   chains with two-body interactions only. In the limit of decoupled chains, each chain displays a transition between a $\mathbb Z_2$-ordered charge density wave (CDW) and a disordered phase \cite{FendleyPRB2004,slagle2021microscopic}. The 1D critical point is described by the Ising conformal field theory (CFT). %, where domain walls  are mapped to Majorana fermion excitations. 
  We consider the regime in which the leading interchain interaction occurs at the  crossings between perpendicular chains, and its strength can be controlled by varying the  layer separation. We first show that the ordered phase of the 2D array retains a \new{subextensive ground state degeneracy inherited from the spontaneous symmetry-breaking of the individual chains.} We refer to this phase as the fractonic CDW (fCDW).
 We analyze the transition from the fCDW to the disordered phase in terms of an array of Ising CFTs coupled at an extensive number of crossings. Using a RG  approach, we find that the interlayer interaction in the effective (2+1)-dimensionally theory behaves as a marginally irrelevant perturbation. As a result,  the critical system displays 1D-like correlations. To confirm the nature of the critical point, we work out   a  Majorana mean-field theory \new{ for an effective lattice model.}  In this approach, the stability of the fixed point of decoupled chains is signalled by the absence of hybridization between  Majorana modes in different chains at weak interchain coupling.   

 This paper is organized as follows. In Sec. \ref{sec:model}, we present the lattice model and discuss the control parameters in the proposed setup. In Sec. \ref{sec:phases}, we analyze the limiting cases of the model, describing  the ground state degeneracy and elementary excitations that characterize  the fCDW phase. In Sec.  \ref{sec:continuum} we take the continuum limit and obtain the effective field theory for the fCDW-disorder transition. The Majorana mean-field theory that supports the conclusion of a  decoupled-chain  critical  point is presented in Sec. \ref{sec:MFT}. We draw some conclusions in Sec. \ref{sec:conclusion}. 
 Finally,  details about \new{the emergent symmetries}  in the continuum limit, the perturbative RG approach and the self-consistent mean-field equations can be found in the appendices.

\section{The model\label{sec:model}}
We consider a model for $N$ trapped Rydberg neutral atoms, which in general are described by the   Hamiltonian \cite{saffman2010quantum,BrowaeysNP2020}
 \be
 H = \sum_{i=1}^N\left[\frac{\Omega}{2}(b_i + b^\dagger_i) -\Delta n_i\right] +  \sum_{1 \leq i < j \leq N} V_{ij}n_i n_j\;,
 \ee
where $b_i, b^\dagger_i$ are   annihilation and creation operators for hardcore bosons for the $i$-th atom, describing the ground state $|g\rangle_i$ and the Rydberg state \new{$|r\rangle_i =b^\dagger_i |g\rangle_i$}, with $n_i \equiv b^\dagger_i b^{\phantom\dagger}_i$. These states are coupled by external lasers with a Rabi frequency $\Omega$ and  a detuning $\Delta>0$. The   van der Waals interaction decays with the distance $R_{ij}$ between atoms as $V_{ij} = C_6 R_{ij}^{-6}$, with a coefficient $C_6>0$. 

 \begin{figure}[t]
     \centering
\includegraphics[width=\columnwidth]{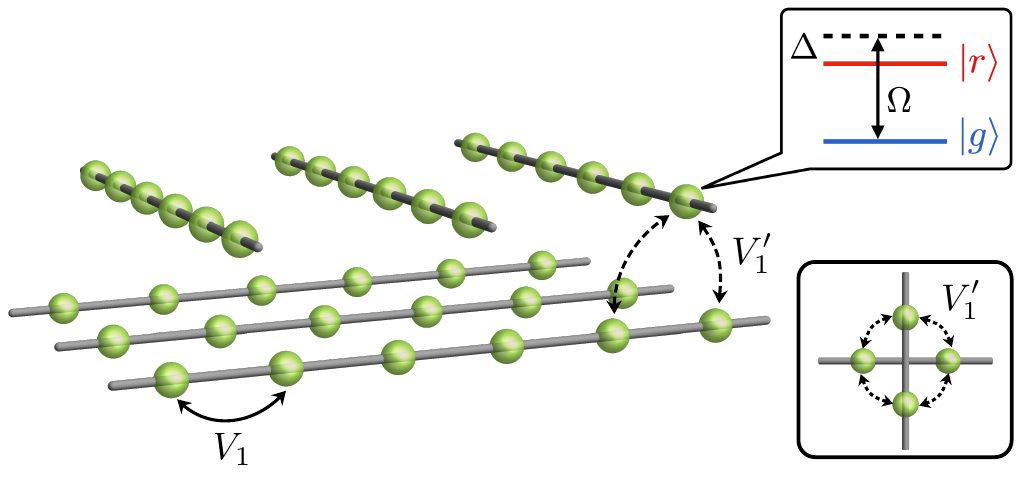}
     \caption{Setup with    Rydberg atoms placed in  chains that belong to two different planes. The atoms are toggled between the ground state $\left|g\right\rangle$ and a Rydberg state $\left|r\right\rangle$ by the external Rabi frequency $\Omega$ with a detuning $\Delta$. The   leading interactions  are the nearest-neighbor intrachain coupling $V_1$  and the interchain coupling   $V_1^\prime$   around a crossing. The lower-right panel shows the top view of a single  crossing. }
     \label{fig:1}
 \end{figure}

The geometry of the lattice determines the leading interactions.  In the setup  of  Fig. \ref{fig:1},  the atoms are placed in 1D chains with lattice spacing $a$. Adjacent parallel chains are separated by a distance $d_\parallel =\nu a$, with an integer $\nu \geq2$, and the layers are separated by $d_\perp$. The shortest distance between perpendicular chains occurs at a ``crossing'' (as viewed from above) where each atom is coupled symmetrically to a pair of atoms in the other layer. This model respects a $C_4$ lattice rotation symmetry, which also exchanges the layers,  and a $\mathbb Z_2$ time-reversal symmetry defined as complex conjugation. Due to the fast decay of the interactions, we consider only two terms, $V_1$ and $V^\prime_1$, corresponding to nearest-neighbor intrachain and interchain couplings, respectively. The ratio $V^\prime_1/V_1 = 8[1+2(d_\perp/a)^2]^{-3}$ varies rapidly with the layer separation. Importantly, we neglect direct couplings between parallel chains.  In Fig. \ref{fig:1} and throughout this work, we represent the array with $\nu=2$ atoms between  two crossings, but in practice it may   be convenient to take $\nu>2$ to further suppress the interaction across the distance $d_\parallel$.
 
 We can map the Hamiltonian describing the interacting Rydberg atoms to a spin model by introducing the  Pauli operators   \be
Z_i=2n_i-1, \qquad X_i=b_i^{\phantom\dagger}+b_i^\dagger.\ee 
In particular, in the limit $d_\perp\to\infty$ we can set $V_1'=0$, and the corresponding Hamiltonian $H_0$ is equivalent to   Ising chains with transverse and longitudinal fields \cite{slagle2021microscopic}:
 \be
H_0= \sum_{\ell=1}^{L_x + L_y}\sum_{m=1}^{ L_\ell}(J Z_{m, \ell}Z_{m+1,\ell} + h_X X_{m, \ell}+h_Z Z_{m, \ell}),\label{H0}
 \ee
 where $J=V_1/4>0$, $h_X=\Omega/2$ and $h_Z=(V_1-\Delta)/2$.  Here we have introduced a notation which is convenient for an array of spin chains: $\ell $ is a chain index, $m$ labels the position along the chain, and  $L_x$ ($L_y$) is the number of vertical (horizontal) chains in the upper (lower) layer. To count the number of sites in each chain, we  define   $L_\ell =\nu L_y$ for $1\leq \ell \leq L_x$ and $L_\ell=\nu L_x$ for   $L_x+1\leq \ell \leq L_x+L_y$.  We also  assume periodic boundary conditions.  For finite $d_\perp$, the  Hamiltonian in Eq. (\ref{H0}) is perturbed by  the interchain coupling 
\be
\delta H = J^\prime \sum_\diamond\sum_{\langle i,j\rangle  \in \diamond} Z_i Z_j\;,\label{interchain}
 \ee
where $J^\prime = V^\prime_1/4\geq0$ and  $\diamond$  stands for the bonds  around a crossing; see Fig. \ref{fig:1}. In addition, $V_1'$ renormalizes the longitudinal field by  $\delta h_Z = V^\prime_1/2$. Hereafter we will discuss the model in terms of the spin variables and  the  parameters $h_X$, $h_Z$, $J$ and  $J'$.

\section{Phase diagram   \label{sec:phases}}
First, let us discuss the limit of decoupled chains $J'=0$, obtained by taking $d_\perp\to\infty$. The phase diagram of a single Ising chain has been studied numerically \cite{ovchinnikov2003antiferromagnetic}. For sufficiently   large $h_X$ or $h_Z$, the system is in a trivial disordered phase.  For $h_X,h_Z\ll J$,  each chain locks into one of   two CDW states represented by $|\cdots rgrg\cdots\rangle$ or $|\cdots grgr\cdots \rangle$, breaking translational invariance. In particular, for $h_Z=0$ we recover the exactly solvable transverse-field Ising chain, for which the  critical point is well known to occur  at $h_X/J=1$. More generally, a critical value of $h_X/J$ exists provided that $|h_Z|<J$. In the ordered phase,  each chain contributes with two states to the ground state degeneracy, which implies  a $2^{L_x+L_y}$-degenerate ground state manifold in this limit. 

   \begin{figure}[t]
     \centering
\includegraphics[width=1.0\columnwidth]{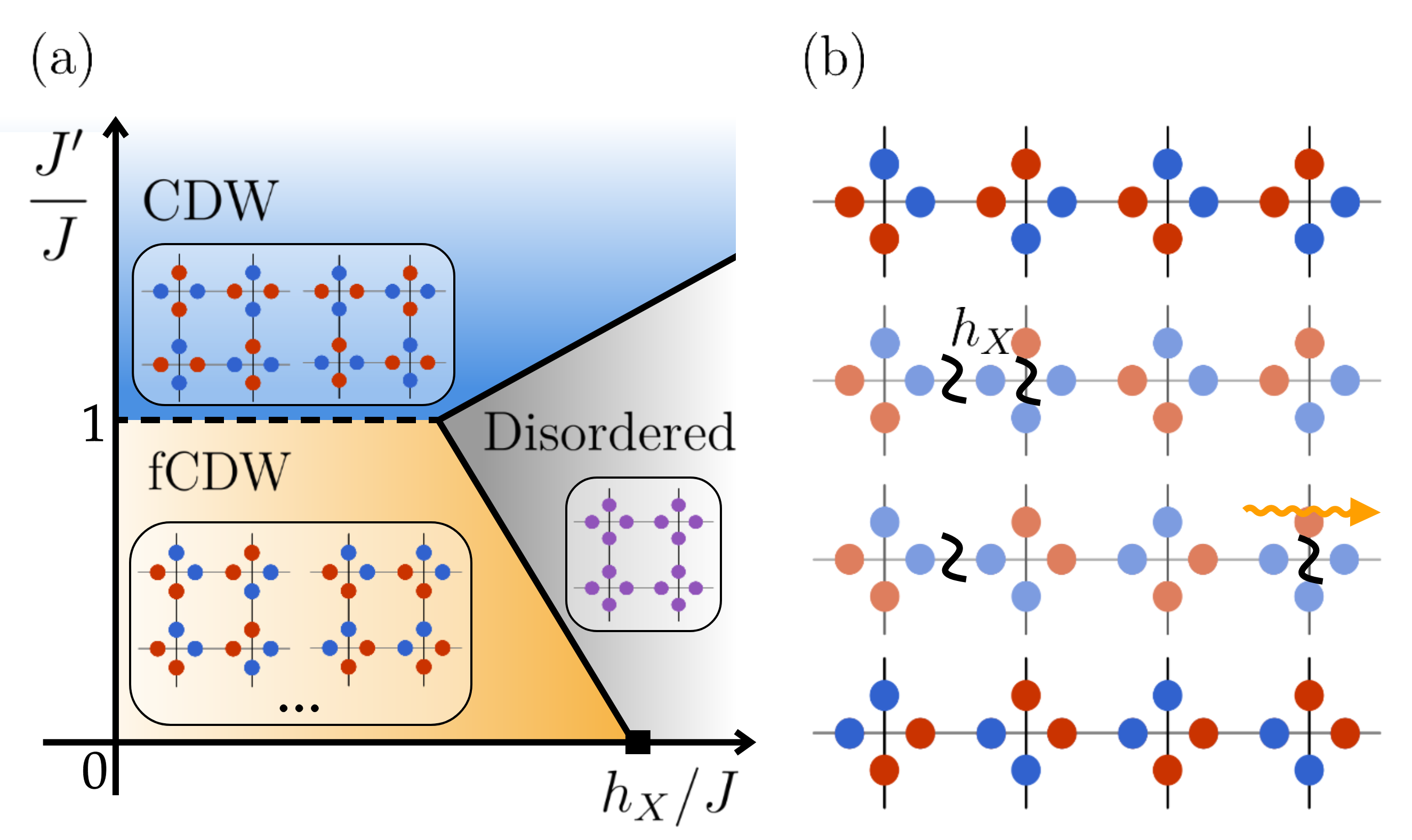}
     \caption{(a) Schematic  phase diagram   for a fixed value of $h_Z\ll J$. The local atomic states $\left|g\right\rangle$ and $\left|r\right\rangle$ in the ordered phases are represented by  blue and red, respectively,   while the  disordered state  is depicted in purple. The square on the $J'=0$ axis marks the critical point of decoupled chains. Solid (dashed) lines denote second-order (first-order)   transitions.   (b) Domain-wall excitations in the fCDW phase. Given an fCDW  ground state (top), reaching another ground state (bottom) requires creating and moving  domain walls around the system.}
     \label{fig:2}
 \end{figure}

To see the difference  from a trivial stack of  1D   states,  we now turn on the interchain  coupling    $J^\prime >0$. In the following we shall  assume that $h_Z$ is fixed and discuss the phase diagram of the 2D model as a function of $h_X/J$ and $J'/J$. 

For $h_X=0$, the Hamiltonian $H = H_0 + \delta H$ reduces to a classical Ising model. For  $h_X=J'=0$, we have  the $2^{Lx+L_y}$ classical   ground states with energy $E_{\rm cl}^{(1)}=-2\nu JL_xL_y$. Note that this energy does not depend on $J'$ due to the frustration of the interchain coupling.  On the other hand, for $h_X=0$ and $J'\gg J$, the classical ground state is only twofold degenerate, corresponding to CDW states in which  the  four atomic states around each crossing alternate as $|  rgrg \rangle$ or $|  grgr \rangle$; see  Fig. \ref{fig:2}(a). These states have energy $E_{\rm cl}^{(2)}=-[4J'+2(\nu-2)J]L_xL_y$. Therefore, increasing $J'$ along the classical line in the phase diagram we encounter a level crossing   at $J'=J$, associated with  a first-order transition between the two types of ordered states.  

One can go from the  twofold-degenerate CDW to the disordered phase by increasing $h_X$. In the strong-coupling limit    $J'/ J\gg 1$, we can project the model onto the eigenstates $\left|\tau^z=1\right\rangle \equiv|  rgrg \rangle$ and $\left|\tau^z=-1\right\rangle \equiv|  grgr \rangle$ of the interaction at each crossing. It is straightfoward to show that the effective Hamiltonian in this limit is an Ising model (on the square lattice for $\nu=2$) with transverse field $\tilde h_X\sim h_X^4/(J')^3$ at the projected sites. As expected from the spontaneous breaking of a global $\mathbb Z_2$ symmetry,  the transition from the CDW to the disordered phase  belongs to the 3D Ising universality class, as found in other  models of Rydberg arrays \cite{samajdar2020complex,yang2022density}.

By contrast,  the ordered phase at   $J'<J$ has a subextensive ground state degeneracy which is robust against quantum fluctuations induced by a weak transverse field. To see this, note that the low-lying gapped excitations in this regime are domain walls created in  pairs by applying a string operator on a given chain $\ell$, \be
S_{\ell}(m_1,m_2)=\prod_{m_1\leq m< m_2} X_{m,\ell},\ee
where $m_1$ and $m_2$ are  the ends of the string. Two states in the ground state manifold can  be coupled only by nonlocal processes that move domain walls around the system, described by  closed strings $S_\ell(m_1,m_1+L_\ell)$ with length $L_\ell$.  The action of this operator corresponds to ``sliding'' the spin   configuration along the chain direction;  see Fig. \ref{fig:2}(b). Moreover, the domain walls in the subspace of low-lying excited states behave as lineons, as their motion is restricted to  the chain direction, with  dispersion \be
E_{\rm dw}(k)=2J -2h_X\cos(ka)+\mc O(h_X^2).\ee 
The motion  of domain walls  discussed here  is   reminiscent of the sliding transformation in the quantum Hall smectic phase \cite{LawlerPRB2004}.

Formally, the 1D  nature of the excitation spectrum can be linked to an emergent symmetry \new{inherited from the decoupled chains}.  Consider the action of a translation in the $\ell$-th chain, \be
\mc T_\ell: \; Z_{m,\ell} \mapsto Z_{m+1,\ell},\quad  X_{m,\ell} \mapsto X_{m+1,\ell}.\ee 
This is not an exact symmetry since it does not commute with the Hamiltonian in the presence of interchain couplings. However, an emergent symmetry can be defined by the condition \cite{nussinov2015compass}
  \be
    [H, \mc P \mc T_\ell\mc  P]=0,
 \ee
 where $\mc P$ is a projector onto a low-energy subspace. Both the ground-state and two-domain-wall subspaces  are stabilized under the action of the group generated by $\mc T_\ell$. This means that the faithful symmetry action at low energies is given by $\mathbb Z_2^{L_x}\times \mathbb Z_2^{L_y}$, yielding the $2^{L_x+L_y}$ degenerate ground states. This exponential  dependence on the linear size   is characteristic of fracton models \cite{Nandkishore2019,pretko2020fracton}, which motivates us to call this  the fCDW phase.   This  symmetry argument   holds provided that   the domain wall pairs are gapped and the low-energy subspaces are clearly separated from multi-particle continua. However,  the domain walls eventually condense as we increase  the transverse field, driving a transition to the disordered phase. In the following we would  like to understand the fate of the \new{emergent symmetry} near this critical point. 
  
 \section{Effective field theory\label{sec:continuum}}

We now move to construct and analyze an effective theory for the transition between the fCDW and the disordered phase. 

 \subsection{Continuum limit}
 
 We begin by putting all the chains at criticality in the uncoupled regime, i.e., near the point represented by a square in Fig. \ref{fig:2}(a).   Each critical chain is described by an Ising CFT  \cite{slagle2021microscopic}. This theory contains two classes of nontrivial primary operators: the energy   operator $\varepsilon$ with   conformal dimensions $\left(\frac 1 2, \frac 1 2\right)$ and the spin field operator $\sigma$, with dimensions $\left(\frac{1}{16}, \frac{1}{16}\right)$ \cite{francesco2012conformal}. Lattice operators can   be expanded as
 \bea   
    X_{m,\ell} &\sim& \langle X_{m,\ell}\rangle \mathbb I + c_\sigma^X (-1)^m \sigma_\ell(x)+c_\varepsilon^X\varepsilon_{\ell}(x) + \cdots\;,\label{Xexpansion}\\
    Z_{m,\ell} &\sim& \langle Z_{m,\ell}\rangle \mathbb I + c_\sigma^Z (-1)^m \sigma_\ell(x)+c_\varepsilon^Z\varepsilon_{\ell}(x) + \cdots\;,\label{Zexpansion}
 \eea
 where $\mathbb I $ is the identity,  $x=ma$ is the position along the chain,  $c_\sigma^{X,Z}$ and $c_\varepsilon^{X,Z}$ are nonuniversal real constants, and we omit higher-dimension operators. The $\mathbb Z_2$  symmetry of the Ising CFT corresponds to the 1D translation, under which the spin field $\sigma_\ell$ changes sign. 
 
The low-energy Hamiltonian for decoupled chains can be written in terms of   Majorana fermions. For a single chain, the  holomorphic and anti-holomorphic parts of the stress tensor are  given by \be
T(x) = \frac{i}2\eta \partial_x \eta,\qquad \bar T (x)= -\frac{i}2\bar \eta \partial_x \bar \eta,\ee
where $\eta(x)$ and $\bar \eta(x)$ are chiral Majorana fermions. The 1D Hamiltonian for a single chain near  criticality is   \be
H_\mathrm{1d} = \int \mathrm dx \; [v(T + \bar T)+m\varepsilon],\ee 
where $v$ is the spin velocity and  the energy operator   $\varepsilon(x) = i\bar \eta \eta(x)$ appears in the mass term. Tuning to the critical point,   we set  $m=0$. Summing over all chains, we have
\be
H_0 \simeq \sum_{\lambda \in \{\text{h,v}\}}\sum_{\ell_\lambda=1}^{L_\lambda} \int \mathrm{d}x_\lambda \;v(T_{\lambda,\ell_\lambda} + \bar T_{\lambda,\ell_\lambda})(x_\lambda) ,
\ee
where we separate the contributions from horizontal (h) and vertical (v) chains by defining the labels $\ell_{\rm h} \in \{1, \cdots, L_{\rm h}\equiv  L_y\}$, $\ell_{\rm v} \in \{1, \cdots, L_{\rm v}\equiv L_x\}$, and the  coordinates $x_{\rm h}=x$ and $x_{\rm v}=y$. 

Next, we add interchain couplings. Note that the   interaction   at a crossing  has the form $J'(Z_{m,\ell}+Z_{m+1,\ell})(Z_{m',\ell'}+Z_{m'+1,\ell'})$. In the continuum limit, this interaction selects the non-oscillating terms in Eq. (\ref{Zexpansion}). We obtain
\begin{equation}
    \delta H =  \frac{v g}{2 \pi} \sum_{k=1}^{N_c} \varepsilon_{\mathrm{h}, \ell_{\mathrm{h}}^k}\left(x_k\right) \varepsilon_{\mathrm{v}, \ell_{\mathrm{v}}^k}\left(y_k\right),\label{discretecrossings}
\end{equation}
where $\mathbf x_k = (x_k,y_k)$ are the positions of the   crossings, which form  a square lattice with spacing $d_\parallel$, the index  $\ell^{k}_{\mathrm h, \mathrm v}$ labels the horizontal and vertical chain which participate in the $k$-th crossing, and $N_c=L_xL_y$ is the total number of crossings. The coupling constant $g$ is of order $J'/J$. We note that the interchain coupling also generates a mass term, but we can tune the mass to zero again by adjusting the longitudinal field.

The Hamiltonian  is invariant under a discrete translational symmetry,  which acts on local operators in  the Ising CFT as
\bea
\mathrm{T}_x&: & O_{\mathrm{v}, \ell}(y) \mapsto O_{\mathrm{v}, \ell+1}(y), \nonumber\\ 
&& O_{\mathrm{h}, \ell}(x) \mapsto O_{\mathrm{h}, \ell}\left(x+d_{\|}\right); \\
\mathrm{T}_y&: & O_{\mathrm{h}, \ell}(x) \mapsto O_{\mathrm{h}, \ell+1}(x), \nonumber \\
&&O_{\mathrm{v}, \ell}(y) \mapsto O_{\mathrm{v}, \ell}\left(y+d_{\|}\right) .
\eea
The interaction in Eq. (\ref{discretecrossings})  perturbs the critical chains with a macroscopic number of defects.
To deal with this interaction, we take a second continuum limit equivalent to sending  $d_\parallel \to 0$, appropriate  when 
the length scales of the observables being probed are considerably larger than $d_\parallel$. The continuum limit of local operators is defined as $O_{\mathrm h, \ell_\mathrm h}(x) \to d_\parallel O_{\mathrm h}(x,d_\parallel \ell_\mathrm h)$ and $O_{\mathrm v, \ell_\mathrm v}(y) \to d_\parallel O_\mathrm v(d_\parallel \ell_\mathrm v,y)$. The sums over the chains are   replaced by integrals:
\begin{equation}
    \sum_{\ell_\lambda=1}^{L_\lambda} (\cdots) \to \int \frac{\mathrm d\bar{x}_\lambda}{d_\parallel} (\cdots)\quad , \quad \lambda \in \{\mathrm h,\mathrm v\},
\end{equation}
where $\bar{x}_\mathrm h= x_\mathrm v = y$ and $\bar{x}_\mathrm v = x_\mathrm h = x$ correspond to the  directions perpendicular to horizontal and vertical  chains, respectively. The effective Hamiltonian is then given by
\be
H \simeq \int \mathrm d^2\mathbf x\,v \left[\sum_{\lambda\in \{\rm h,\rm v\}}(T_\lambda+ \bar T_\lambda)(\mathbf x) + \frac{g}{2\pi} \varepsilon_{\rm h} \varepsilon_{\rm v}(\mathbf x)\right],
\label{eq:rydbergstack}
\ee
where $\mathrm d^2\mathbf x = \mathrm dx\mathrm dy$. 

To properly define a 2D field theory, we need to specify the corresponding operator product expansions. First, let us take the limit $g\to 0$, in which case the fields in Eq. (\ref{eq:rydbergstack}) act as coarse-grained versions of the operators in the Ising CFT. Before we  take the $d_\parallel\to0$ limit,  the two-point  functions of the nontrivial primaries read
\begin{align}
    \langle\sigma_{\lambda, \ell_\lambda}\left(\mathbf{r}_\lambda\right) \sigma_{\lambda, \ell_\lambda^{\prime}}(\mathbf{0})\rangle_0 &=\frac{\delta_{\ell_\lambda, \ell_\lambda^{\prime}}}{\left|\mathbf{r}_\lambda\right|^{1 / 4}} ,\\
    \langle\varepsilon_{\lambda, \ell_\lambda}\left(\mathbf{r}_\lambda\right) \varepsilon_{\lambda, \ell_\lambda^{\prime}}(\mathbf{0})\rangle_0&=\frac{\delta_{\ell_\lambda, \ell_\lambda^{\prime}}}{\left|\mathbf{r}_\lambda\right|^2}  , \quad \lambda \in\{\mathrm{h}, \mathrm{v}\},
\end{align}
 where $\mathbf r_{\mathrm h} = (x, v \tau)$ and $\mathbf r_{\mathrm v} = (y,v\tau)$ are coordinates in the (1+1)-dimensional Euclidean spacetime with imaginary time $\tau$. All correlators of the form $\langle O_\mathrm h O_\mathrm v\rangle_0$ vanish. We then take the second  continuum limit, \new{defining  $\mb R=(x,y,v\tau)$  and  replacing $\delta_{\ell_\lambda, \ell^\prime_\lambda} \to \delta(d_\parallel \ell_{\lambda}-d_\parallel \ell^\prime_{\lambda})$.} Now, the same correlation functions written in the (2+1)-dimensional theory become
 \bea
      \langle \sigma_\lambda(\mathbf R)\sigma_\lambda(\mb 0) \rangle &=&   \frac{f_\lambda(x,y)}{|\mathbf R|^{1/4}}, 
     \\
     \langle \varepsilon_\lambda(\mathbf R) \varepsilon_{\lambda}(\mathbf 0) \rangle_0 &=& \frac{f_\lambda(x,y)}{|\mathbf R|^2}  ,\quad \lambda \in \{\mathrm h, \mathrm v\},\label{correlator}
 \eea
 \new{where  $f_\mathrm h(x,y) = \delta(y)$ and $f_\mathrm v(x,y) = \delta(x)$.}  Similar anisotropic correlators arise in theories of sliding Luttinger liquids \cite{MukhopadhyayPRB2001,StarykhPRL2002}.  To recover the 1D behavior of intrachain correlators, we must regularize the delta function at short distances as $\delta(0)\to \frac1{d_\parallel}$.   Higher-point functions or correlators of other local operators (say, involving  descendants or the stress-energy tensor) can be similarly regularized. This procedure perturbatively defines the 2D theory   in Eq. (\ref{eq:rydbergstack}).
 
\subsection{\new{Emergent symmetries}}
 
The peculiar low-energy Hamiltonian in Eq. (\ref{eq:rydbergstack}) inherits symmetries and dualities from the Ising chains. In the CFT, these symmetries are implemented by topological defect line operators labeled by the primary fields \cite{kramers2004frohlich,chang2019topological}. First, there is a  Kramers-Wannier duality implemented by $D^\sigma=\bigotimes_\ell D_\ell^\sigma$, where $D_\ell^\sigma$ is the $\sigma$ defect of the $\ell$-th chain  \cite{kramers2004frohlich}.  The action of $D^\sigma$ takes $\varepsilon_\ell \mapsto -\varepsilon_\ell$, exchanging  the correlators of the fCDW and the disordered phase, and becomes a global symmetry at the critical point. Second and more interestingly, \new{the emergent  $\mathbb Z_{2}^{L_x}\times \mathbb Z_2^{L_y}$ symmetry} is manifested as the $\varepsilon$ defect, which acts on horizontal chains as \be
D^\varepsilon_{\ell_{\rm h}} \sigma_{\rm h}(x, \ell_{\rm h}a) = -\sigma_{\rm h}(x,\ell_{\rm h}a) D^\varepsilon_{\ell_{\rm h}},\ee 
and similarly for vertical chains. If we allowed for direct couplings between the order parameters of parallel chains, the symmetry would be lowered to a global $\mathbb Z_2$ symmetry. 

The defect line operators  form the following fusion algebra \cite{kramers2004frohlich}:
\bea
&&D^\varepsilon_\ell\times D^\varepsilon_\ell = 1 ,
\label{eq:idZ2}\\
&&D^\varepsilon_\ell\times D^\sigma 
 =D^\sigma\times D^\varepsilon_\ell = D^\sigma, \\
&&D^\sigma\times D^\sigma = \bigotimes_{\ell=1}^{L_x+L_y}(1+ D^\varepsilon_\ell)\;.
\label{eq:idKW}
\eea
Here, $\times$ denotes the fusion,  $\otimes$ takes the tensor product of symmetry lines on different chains, and we omit the trivial fusion rules involving the identity line defect. \new{A similar version of this algebra was recently discussed in Ref. \cite{weiguang2023subsystem} on the lattice, as an example of a non-invertible subsystem symmetry. However, in our case, the  algebra is simpler, descending from symmetries of decoupled chains. It is a non-invertible symmetry since the action of the Kramers-Wannier defect does not admit an inverse. It is known that the existence of such operators strongly constrains the low-energy spectrum \cite{seiberg2024non}: For example, the existence of such line operators imposes that the Hilbert space of a perturbed  1+1 CFT cannot be trivially gapped \cite{chang2019topological}.  This argument can be adapted to our realization, since the algebra comes from a tensor product of CFTs, when the total number of chains $L_x +L _y$ is odd;  see   App. \ref{app:noninvertible}.}  Thus, we have the guarantee that the model defined by Eq. (\ref{eq:rydbergstack}) cannot have a trivial ground state even at strong coupling.

\subsection{Perturbative RG analysis\label{sec:RG}}
We proceed to analyze the model perturbatively at weak coupling $g \ll 1$, corresponding to  $J^\prime \ll J$. The fate of the theory   relies on the RG flow of the effective coupling $g(s)$ at length scale $s$ \cite{cardy1996scaling}. Power counting based on the correlator in Eq. (\ref{correlator})   indicates that the coupling has effective scaling dimension $\Delta=3$ and is marginal at tree level. We calculate the  beta function $\beta (g)$ to leading order using the lattice spacing $a$ as a short-distance cutoff and introducing the large-distance cutoffs $D_x$ and $D_y$ for the $x$ and $y$ directions; see App.  \ref{app:RG}. The leading contribution appears at two-loop level and is strongly affected by the 1D nature of the correlators. We find that $g$ behaves as a marginally irrelevant coupling, with   beta function     
\be
    \frac{dg}{dl}=- \ln \left(\frac{D_xD_y}{a^2}\right)g^3 + \cdots\;,\label{betafunc}
\ee
where $l=\ln(s/a)$  in the regime $a\ll s\ll \text{min}(D_x,D_y)$.  

Remarkably, the beta function   depends explicitly on the ratio between the infrared and ultraviolet cutoffs, as found in  models with UV-IR mixing \cite{GrosvenorPRB2023, kapustin2018wilsonian}. Physically, we expect the scales $D_x$ and $D_y$ to be of the order of the linear system size in the $x$ and $y$ directions, respectively. For $D_{x,y}\sim L_{x,y}a$,   the coefficient in Eq. (\ref{betafunc}) is proportional to $\ln(N_c)$, with  $N_c\gg 1$ being the number of crossings.  Here we adopt the perspective that experiments with Rydberg atoms must be performed on a finite system where  $\ln(N_c)$  is a constant of order 1, but there are notorious subtleties in taking the continuum and thermodynamic limits in the presence of UV-IR mixing; see Refs.  \cite{GorantlaPRB2021,LakePRB2022}.

Since the effective interchain coupling decreases at low energies, correlation functions can be calculated by perturbation theory. In particular, the leading  contribution to correlators of crossed chains comes from the interaction at their crossing. Consider, for instance,  the equal-time   correlator for the energy operator in a pair of horizontal and vertical chains. Taking  a position  $x$ along the $\ell_{\mathrm h}=0$ horizontal chain and position $y$ along the $\ell_{\mathrm v}=0$ chain, we obtain to first order in $g$: 
\bea
 \left\langle\varepsilon_{\mathrm h}(x, 0) \varepsilon_{\mathrm v}(0, y)\right\rangle &\simeq& -\frac{g}{2\pi}\int \mathrm{d}^3\mathbf R\;\langle \varepsilon_{\mathrm h}(x,0) \varepsilon_{\mathrm h}(\mathbf R)\rangle_0\nonumber \\
 &&\times\langle\varepsilon_{\mathrm v}(0,y)\varepsilon_{\mathrm v}(\mathbf R) \rangle_0   .
\eea
Using the intrachain correlator in Eq. (\ref{correlator}) and integrating over the spatial and temporal coordinates, we obtain 
\be
 \langle \varepsilon_{\rm h}(x,0)\varepsilon_{\rm v}(0,y)\rangle \sim -\frac{g}{2(x^2|y|+|x|y^2)} + \mathcal O(g^3)\;.
\label{eq:weakcouplingcorr}
\ee
Note the unusual spatially anisotropic  power-law decay, a prominent feature of fractonic behavior \cite{YouMoessnerPRB2022}. The homogeneous function of degree 3 in the denominator  is consistent with the energy operator  having an effective scaling dimension $\Delta_\varepsilon=3/2$ in the (2+1)-dimensional theory, as expected from  Eq. (\ref{correlator}). However, the correlator is singular for $x\to0$ or $y\to0$, which corresponds to taking two points on the same chain.  When one of the coordinates is of the order of the lattice spacing, say $x\sim a $, we    recover the correlator of the Ising CFT by  $\langle \varepsilon_{\rm h}(a,0)\varepsilon_{\rm v}(0,y)\rangle   \sim    1/y^2$ for $|y|\gg a$.   The result in Eq. (\ref{eq:weakcouplingcorr})  captures the long-distance behavior of the correlation for the    operator $Z_{j,\ell}+Z_{j+1,\ell}$, which can be measured by means of  snapshots of the atomic states in  Rydberg arrays \cite{SemeghiniScience2021,SchollNature2021}.    
 
\section{Majorana mean-field theory\label{sec:MFT}}
The perturbative RG   analysis indicates that the fixed point of decoupled chains, $g=0$, is stable against the interchain coupling.  To  test this picture, we study a lattice model that reduces   to   Eq. (\ref{eq:rydbergstack}) in the continuum limit but also regularizes the short-distance behavior. We consider the effective Hamiltonian 
\be
\tilde H= \sum_{\ell,m}(J_\mathrm e \tilde Z_{m, \ell}\tilde Z_{m+1,\ell} + h_\mathrm e \tilde X_{m, \ell}) + J^\prime_\mathrm e\sum_\diamond\sum_{\langle i,j\rangle  \in \diamond} \tilde X_i \tilde X_j   ,
\label{eq:effhamsubsys}
\ee
where    the parameters $J_\mathrm e$, $h_\mathrm e$ and  $J_\mathrm e ^\prime$ can be chosen so as to tune to the critical point and  to match $v$ and $g$  in the continuum limit.  We denote the new Pauli operators by $\tilde X_i$ and $\tilde Z_i$ to avoid confusion with the original lattice model.  One advantage of Eq. (\ref{eq:effhamsubsys}) over the  original model  is that \new{the 1D symmetry} is now manifest and on-site. This symmetry can be implemented by applying   $\prod_{m=1}^{L_\ell} \tilde X_{m, \ell}$, which takes $\tilde Z_{m, \ell}\mapsto - \tilde Z_{m, \ell}$ for all sites that belong to a given  chain $\ell$. 

Moreover, we can perform a generalized Jordan-Wigner transformation \cite{sachdev2011quantum} and map the intrachain terms of Eq. (\ref{eq:effhamsubsys}) onto a stack of crossed Kitaev chains \cite{kitaev2001unpaired}, in close connection with the Majorana representation in the   field theory.  We introduce two Majorana fermions at each site so  that  \be
\tilde X_{m,\ell}\mapsto i\gamma_{m,\ell}^0\gamma_{m,\ell}^1,\ee
with the anticommutation relation  $\{\gamma^b_{m, \ell} , \gamma^{b'}_{m^\prime, \ell^\prime}\}= 2 \delta^{bb'} \delta_{m, m^\prime} \delta_{\ell, \ell^\prime}$. 
The other components of the local spin operator are written as \bea
\tilde Y_{m, \ell} &=& \eta_\ell B_{m,\ell}\gamma^0_{m,\ell}, \\ 
\tilde Z_{m,\ell} &=& \eta_\ell B_{m,\ell}\gamma^1_{mi,\ell}.
\eea
 To ensure the Pauli algebra of physical operators, we have introduced the  string operators
\be
\quad B_{m,\ell} = \prod_{m^\prime \leq m} i \gamma^0_{m^\prime,\ell}\gamma^1_{m^\prime,\ell}\;,
\ee
and the chain-dependent  Klein factors  $\eta_\ell$ \cite{Crampe2013} that   obey $\{\eta_{\ell}, \eta_{\ell^\prime}\} =2 \delta_{\ell, \ell^\prime}$ and commute with the ``dynamical'' Majorana modes $\gamma^b_{m,\ell}$. The Hamiltonian is written  in terms of Majorana fermions as
\bea
\tilde H &=& \sum_{\ell, m}(iJ_\mathrm e \gamma^1_{m, \ell} \gamma^0_{m+1,\ell} + ih_\mathrm e \gamma^0_{m, \ell}\gamma^1_{m,\ell}) \nonumber\\
&&- J^\prime_\mathrm e\sum_\diamond \sum_{\langle i, j \rangle \in \diamond} \gamma^0_i \gamma^1_i\gamma^0_j \gamma^1_j\;.\label{Majoranamodel}
\eea

\begin{figure}[t]
     \centering
\includegraphics[width=1.0\columnwidth]{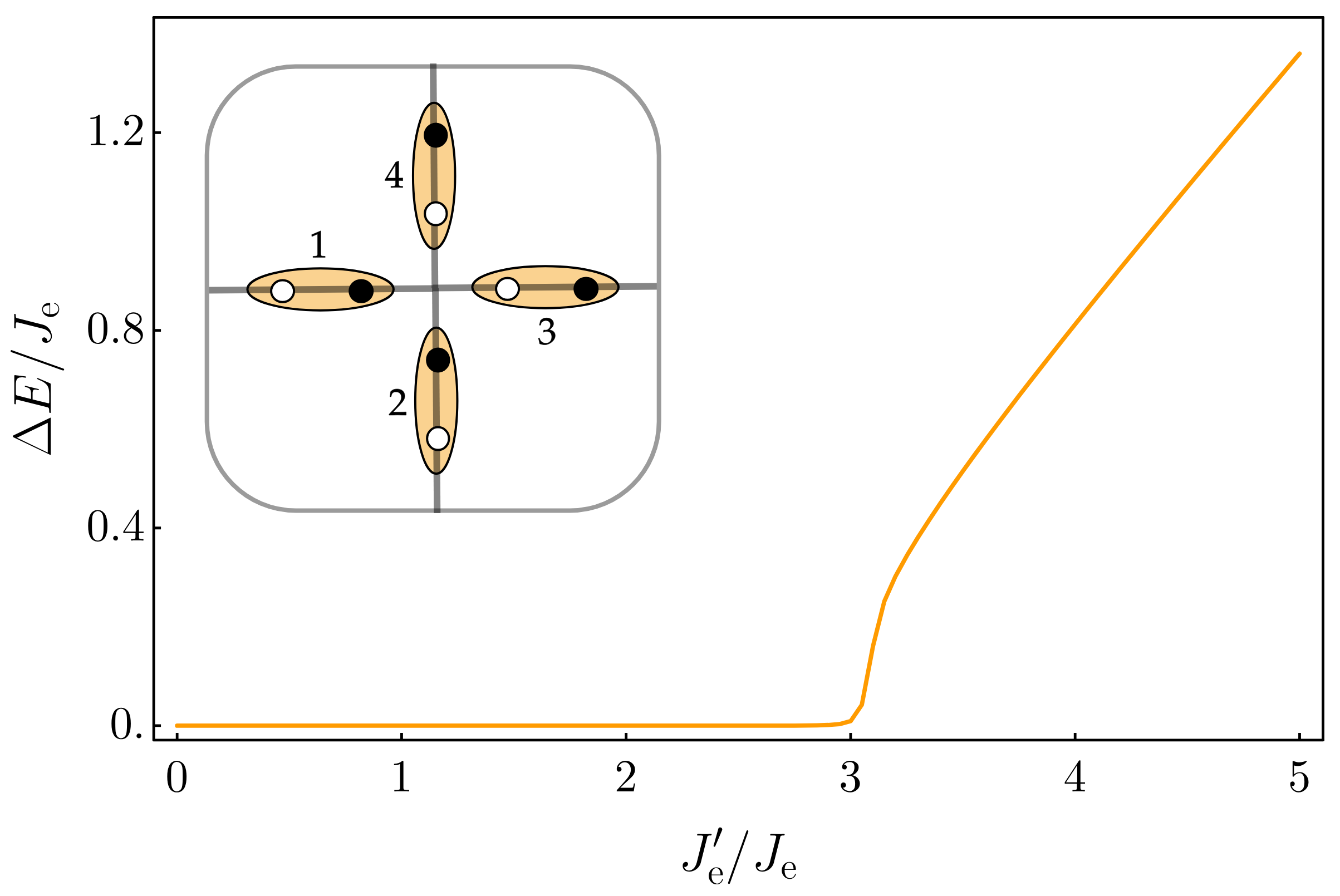}
     \caption{Energy  gap for Majorana fermion excitations calculated  by solving the mean-field equations for the  model in Eqs. (\ref{eq:effhamsubsys}) and (\ref{Majoranamodel}) with $h_{\rm e}=J_{\rm e}$ tuned to the critical point of the Ising chains. Here we focus on the lattice with $\nu=2$. The vanishing gap at weak coupling is expected for a fixed point of decoupled chains.   Inset: unit cell with eight  Majorana modes $\gamma^a_{\mb R,\alpha}$. Each  site $\alpha \in \{1,2,3,4\}$  contains  two modes,  $b \in\{0,1\}$, represented by white and black dots, respectively.}
     \label{fig:3}
 \end{figure}

The unit cell for the effective Majorana model on  the $\nu=2$ lattice is shown in the inset of Fig. \ref{fig:3}. We denote the  8 Majoranas within each unit cell  by $\gamma^b_{\mb R,\alpha}$, where   $b\in\{0,1\}$, $\mb R$ is the position of the unit cell, and $\alpha\in \{1,2,3,4\}$ labels the sites around the crossing. The symmetries act projectively due to the gauged fermion parity. Time reversal    conjugates complex numbers and   takes  $\gamma^b_{\mb R,\alpha} \mapsto (-1)^{b+1}\gamma^b_{\mb R,\alpha}$, while the $C_4$ symmetry  acts as $\gamma^b_{\mb R,\alpha} \mapsto \gamma^{b'}_{\mb R^\prime,\alpha+1}$, where $b'=b+\cos^2(\frac{\pi \alpha}2) \text{ (mod 2)}$   and $\mb R'$ is the rotated position. 

We treat the quartic  interaction in the fermionic representation of Eq. (\ref{Majoranamodel}) using a Majorana mean-field approach \cite{Rahmani2019,ChenPRB2019,SlaglePRB2022}. In this approach, a departure from the decoupled-chain fixed point is signalled by a spontaneous hybridization between modes in perpendicular chains. We assume that the mean-field ansatz at criticality  respects time reversal invariance because this symmetry is preserved both in the fCDW and in the disordered phase.  Imposing  time-reversal as well as translation invariance, we obtain eight mean-field parameters allowed by symmetry: 
\be
A_\alpha = \langle i \gamma^0_{\mb R,\alpha}\gamma^1_{\mb R, \alpha+1}\rangle \; ,\quad B_\alpha = \langle i\gamma^1_{\mb R, \alpha}\gamma^0_{\mb R, \alpha+1}\rangle\;.
\label{eq:mftparams}
\ee 
Note that the mean-field decoupling of the interaction  also generates the on-site  amplitudes $ \langle i \gamma^0_{\mb R, \alpha}\gamma^{1}_{\mb R, \alpha}\rangle$, but the latter can  be absorbed into a renormalization of the transverse field $h_{\rm e}$, which must be tuned to the critical point.

We diagonalize the quadratic mean-field Hamiltonian and solve the   self-consistency  equations numerically; see App.  \ref{app:MFT} for details. For $h_\mathrm e= J_\mathrm e$ and small $J_{\rm e}'$, we find that    both $A_\alpha$ and $B_\alpha$ vanish and the fermionic spectrum is equivalent to critical Kitaev chains. As a consequence, the Majorana fermions are restricted to move within the respective chains. As we increase $J_{\rm e}'$, the hybridization parameters eventually become nonzero and the 2D system develops an energy gap; see Fig. \ref{fig:3}. In this regime, the $C_4$ symmetry is spontaneously broken. Note, however, that the gapped regime   occurs at strong coupling, $J_{\rm e}'>J_{\rm e}$, where the connection with the original model via the effective field theory in Eq. (\ref{eq:rydbergstack}) breaks down. While we cannot rule out additional phases around the tricritical point in Fig. \ref{fig:2}(a), the mean-field theory confirms  that the transition  at weak to intermediate coupling is governed by the decoupled-chain fixed point. As characteristic of subdimensional criticality, at this fixed point we obtain  \new{further emergent 1D symmetries. In the effective field theory with $g\to0$, the  symmetry associated with the Kramers-Wannier defect $D^\sigma$ is enlarged, and the resulting symmetries are generated by    \emph{both} $D_\ell^\sigma$ and $D_{\ell}^\varepsilon$  for each chain.}

 \section{Conclusions \label{sec:conclusion}}
 We proposed  a model for a fractonic quantum phase transition in Rydberg arrays. The setup consists of two layers of Rydberg chains in which  the dominant interchain coupling occurs between pairs of perpendicular chains. If we neglect interactions between parallel chains, the ordered phase at weak coupling corresponds to a fCDW phase whose ground state degeneracy increases exponentially with the number of chains or with the linear system size.
 
 Starting from the fCDW phase and increasing the quantum fluctuations,  we cross a  transition to a disordered phase.  We  studied the critical  point using an effective field theory  in 2+1 dimensions that inherits properties of the Ising CFT. We also  constructed a Majorana mean-field approach for a lattice model that reduces to the same field theory in the continuum limit. Our analysis shows that the critical point exhibits  particles with restricted mobility,  \new{emergent symmetries},   and  anisotropic correlators that manifest the UV-IR mixing.
 
 Moving forward, it would be interesting to explore   non-equilibrium dynamics near criticality,  extensions to   $\mathbb Z_n$-ordered phases \cite{FendleyPRB2004}, and to apply  numerical methods  \cite{Samajdar2021,merali2023stochastic} to study  the fCDW phase and the associated transitions. Our work represents a significant step towards the realization of fracton-like physics in quantum simulation platforms.

\begin{acknowledgments}
We acknowledge funding by  Brazilian agencies Coordena\c{c}\~ao de Aperfei\c{c}oamento de Pessoal de N\'ivel Superior  (R.A.M.) and Conselho Nacional de Desenvolvimento Cient\'ifico e Tecnol\'ogico (R.G.P.). This work was supported by a grant from the Simons
Foundation (Grant No. 1023171, R.G.P.). Research at IIP-UFRN is supported by Brazilian ministries MEC and MCTI. 
\end{acknowledgments}

\appendix

\section{Non-invertible emergent symmetry and nontriviality of the ground state \label{app:noninvertible}}

We will argue that   the non-invertible symmetry implies a nontrivial ground state for an odd number of chains. As discussed in the main text, there is a $\mathbb Z_2^{L_x}\times \mathbb Z_2^{L_y}$ symmetry generated by the line operators $\{D^\varepsilon_\ell\}_{\ell=1}^{L_x+L_y}$ and a non-invertible symmetry generated by a Kramers-Wannier line $D^\sigma$ at low energies for $g\neq 0$. We note that the fusion rules in Eqs. (\ref{eq:idZ2})-(\ref{eq:idKW}) depend on the number of chains, and therefore are not well defined in the thermodynamic limit.
Similar behavior has been observed   in the construction of the Kramers-Wannier defect (and the corresponding algebra with the $\mathbb Z_2$ symmetry) in the transverse-field Ising chain \cite{seiberg2024non} and in a related lattice model   \cite{weiguang2023subsystem}.

\begin{figure}[t]
    \centering
    \includegraphics[width=0.95\columnwidth]{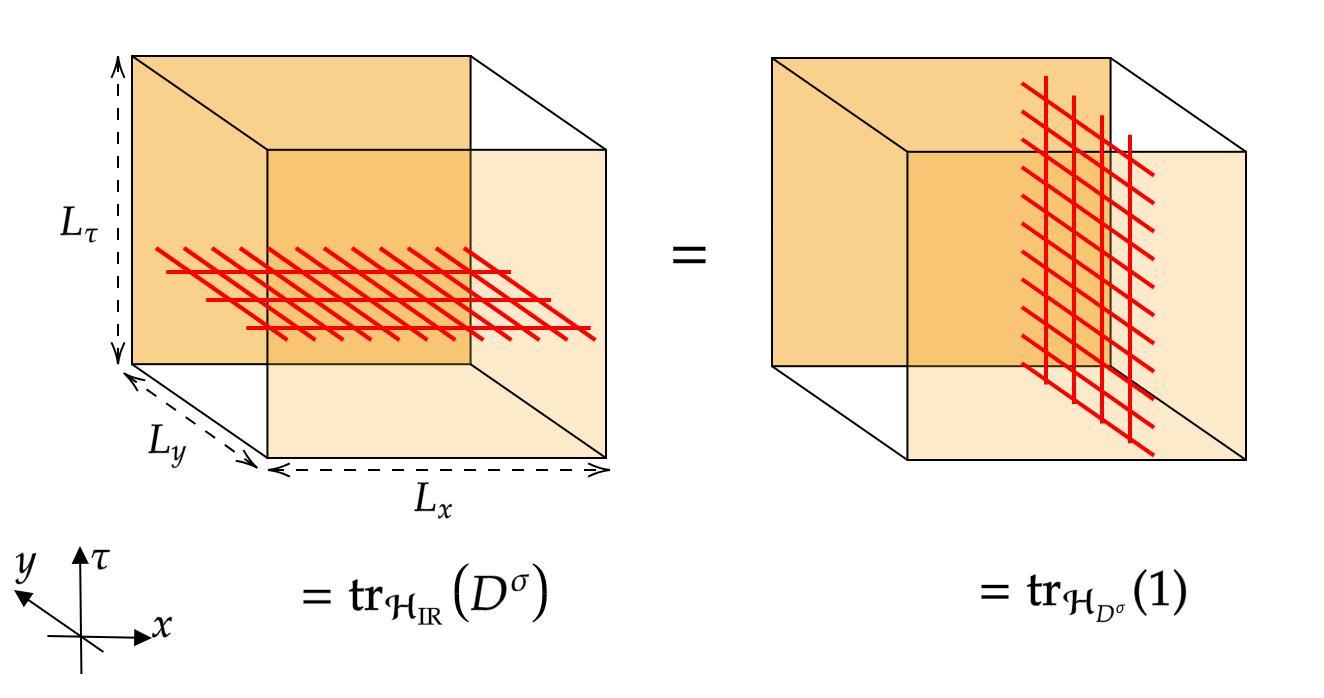}
    \caption{Two ways of inserting the Kramers-Wannier defect $D^\sigma$ in the path integral. On the left, the defect is defined on a constant-time slice, and  its expectation value is computed in the infrared state. On the right, the defect is placed along the time direction, on a constant-$x$ slice. In this case, $L_\tau$ is taken such that $L_\tau = L_x$ to  compute the dimension of the Hilbert space of the defect. At low energies, the results should match.}
    \label{fig:defectbox}
\end{figure}

The expectation values of line operators are invariant under the RG flow. The reason is as follows. Given  a vacuum $|\Omega\rangle$, we define the quantum dimensions \bea
d^\sigma &\equiv& \langle D^\sigma\rangle =\langle \Omega |D^\sigma|\Omega\rangle,\\
d^\varepsilon_S &\equiv& \langle \otimes_{\ell\in S}D^\varepsilon_\ell\rangle = \langle \Omega |\otimes_{\ell \in S} D^\varepsilon_\ell |\Omega\rangle ,\eea
for any subset  $S$ of the total set of chains. It follows from   the fusion ring  that   $\{d^\sigma, \{d^\varepsilon_S\}\}$ satisfy polynomial equations with integer coefficients. Therefore, they must  be RG invariants. This is completely analogous to the argument made in Ref. \cite{chang2019topological} in the context of $(1+1)$-dimensional QFTs.

First, for $g=0$, consider the $(\mathrm{Ising})^{\otimes(L_x +L_y)}$ theory. By radial quantization, the ground state of this theory is $|\Omega_0\rangle\equiv \otimes_{\ell=1}^{L_x+L_y}|1\rangle_\ell$, where $|1\rangle_\ell$ is the state corresponding to the identity operator of the $\ell$-th chain. It is also known that the Kramers-Wannier defect has a quantum dimension of $\sqrt 2$ since $D^\sigma_\ell |1\rangle_\ell = \sqrt 2 |1\rangle_\ell$. Therefore,
\be
D^\sigma |\Omega_0\rangle \equiv \langle D^\sigma \rangle_0|\Omega_0\rangle = 2^{(L_x+L_y)/2}|\Omega_0\rangle\;.
\ee
Thus, even for $g\neq 0$, we have $d^\sigma  = \langle D^\sigma \rangle = \langle D^\sigma\rangle_0=2^{(L_x+L_y)/2}$ by the arguments above.

Suppose now that there is a unique ground state in the infrared described by a Hilbert space $\mathcal H_\mathrm{IR}$ such that $\mathrm{dim}(\mathcal H_\mathrm{IR})=1$. In this case, 
\be
d^\sigma = \langle D^\sigma \rangle =\mathrm{tr}_{\mathcal H_\mathrm{IR}} (D^\sigma) .
\ee
Physically, we can understand this equation in the path integral approach where the trace corresponds to taking periodic boundary conditions in the time direction. Thus, $d^\sigma$ computes the expectation value of $D^\sigma$ on a time slice. 
The same amplitude can be computed by inserting the mesh in the time direction. In the path integral picture, this works by slicing the Euclidean time direction into either $T = \Delta_x \tau L_x$ or $T = \Delta_y\tau L_y$, depending on whether the mesh is positioned in the $x$ or $y$ direction, making the action of $D^\sigma$ well defined; see Fig. \ref{fig:defectbox}.

As a consequence, one can interpret the quantization of  $d^\sigma$ as counting the dimension of a twisted Hilbert space $\mathcal H_{D^\sigma}$, where the operators (and corresponding states) have twisted boundary conditions with the action of $D^\sigma$. But this is the same as computing the corresponding quantum dimension:
\be
d^\sigma = \langle D^\sigma \rangle =\mathrm{tr}_{\mathcal H_{D^\sigma}}(1)  \;,
\ee
leading to a contradiction for $L_x+L_y$ odd, since the rhs is a non-negative integer and the lhs is not. Therefore, the infrared Hilbert space must have more than one state.

\section{Derivation of the RG equation\label{app:RG}}

We now derive the beta function for the coupling constant in the (2+1)-dimensional theory. Consider the partition function associated with  the Hamiltonian in  Eq. (\ref{eq:rydbergstack}). The   path integral can be written in terms of the free partition function $Z_0$, defined at $g=0$, as
\bea
\frac{Z}{Z_0} &=& \left\langle e^{- \frac{g}{2\pi}\int d^3\mathbf R\; \varepsilon_{\mathrm h}\varepsilon_{\mathrm v}(\mathbf R)}\right\rangle_0 \nonumber\\
&=&1-g I^{(1)} + \frac{g^2}{2!}I^{(2)} -\frac{g^3}{3!}I^{(3)} +\cdots\;,
\label{eq:partitionfunc}
\eea
where $d^3\mb R=dxdyd(v\tau)$ is the volume element  in Euclidean spacetime, $\langle\cdots\rangle_0$ denotes the expectation value in the free theory,  and in the second line we expressed the ratio in a perturbative expansion.  The corresponding integrals up to the third order are given by 
\bea
I^{(1)}&=& \frac1{2\pi} \int d^3\mathbf R \;\langle \varepsilon_{\mathrm h} \varepsilon_{\mathrm v}(\mathbf R)\rangle_0\;,\nonumber\\
I^{(2)}&=& \frac1{(2\pi)^2}\int  d^3\mathbf R_1d^3\mathbf R_2 \;\langle \varepsilon_{\mathrm h} \varepsilon_{\mathrm v}(\mathbf R_1)\varepsilon_{\mathrm h} \varepsilon_{\mathrm v}(\mathbf R_2)\rangle_0\;,\nonumber\\
I^{(3)}&=& \frac1{(2\pi)^3}\int d^3\mathbf R_1d^3\mathbf R_2 d^3\mathbf R_3\;\nonumber \\
&&\times \langle \varepsilon_{\mathrm h} \varepsilon_{\mathrm v}(\mathbf R_1)\varepsilon_{\mathrm h} \varepsilon_{\mathrm v}(\mathbf R_2)\varepsilon_{\mathrm h} \varepsilon_{\mathrm v}(\mathbf R_3)\rangle_0 \label{eq:thirdordercont}. 
\eea

The integrals must be regularized  by imposing a UV  cutoff. We   choose a specific cutoff scheme following the approach explained in  Ref. \cite{cardy1996scaling}. The three steps behind the perturbative renormalization group are: (1) perform an infinitesimal RG transformation, where the short-distance cutoff $a$ is renormalized as $a \to a(1+ dl)$, (2) discard all contributions which are $\mathcal O(dl^2)$ or higher; (3) impose that the partition function must remain invariant, reading off the corresponding renormalization conditions. This is the standard approach in the study of scale-invariant fixed points, but we will encounter   difficulties related to UV-IR mixing   when we perturb around the  fixed point of decoupled crossed chains. 

The first-order term is invariant under the scaling, implying that the perturbation is tree-level marginal as mentioned in the main text. The second-order term only contributes to  the renormalization of irrelevant interactions, such as $\varepsilon_{\mathrm h} \partial_x \varepsilon_{\mathrm h}$ and $\varepsilon_{\mathrm v} \partial_x \varepsilon_{\mathrm v}$. The renormalization of the coupling appears   at third order or two-loop level. By Wick's theorem, the contribution in Eq. (\ref{eq:thirdordercont}) can be written as
\bea
 I^{(3)}&=&\frac3{(2\pi)^3}\int d^3\mathbf R_1d^3\mathbf R_2 d^3\mathbf R_3\; \langle \varepsilon_\mathrm h(\mathbf R_1) \varepsilon_\mathrm v(\mathbf R_3)\rangle_0  \nonumber\\
 && \times\langle \varepsilon_\mathrm v(\mathbf R_1) \varepsilon_\mathrm v(\mathbf R_2)\rangle_0 \langle \varepsilon_\mathrm h(\mathbf R_2) \varepsilon_\mathrm h(\mathbf R_3)\rangle_0 \nonumber\\
 &&+ (\mathrm h \leftrightarrow \mathrm v) ,
\eea
where the combinatorial factor comes from exchanging the positions of the interaction vertices. Taking the leading contributions as $\mb R_3\to \mb R_1$ and using the correlator for decoupled chains in Eq. (\ref{correlator}), we obtain
\bea
 I^{(3)}&=&\frac6{(2\pi)^3}\int d^3\mathbf R\, \langle \varepsilon_{\mathrm h}\varepsilon_{\mathrm v}(\mathbf R)\rangle_0\nonumber\\
&&\times \int \frac{dy_{12}d(v\tau_{12})dx_{23} d(v\tau_{23})}{[y_{12}^2+(v\tau_{12})^2][x_{23}^2+(v\tau_{23})^2]},
\eea
where $\tau_{12}=\tau_1-\tau_2$, $\tau_{23}=\tau_2-\tau_3$, $x_{23}=x_2-x_3$ and $y_{12}=y_1-y_2$.
This is the point where the calculation departs from the standard scheme for conformally invariant fixed points. Note the anisotropic dependence of the integrand in the  four-dimensional space spanned by $(x_{12},y_{23},\tau_{12},\tau_{23})$. Since  the integrand is singular for $y_{12}=\tau_{12}=0$ or $x_{23}=\tau_{23}=0$,  we cannot simply integrate out a spherical shell in the four-dimensional space. This dependence can be traced back to the 1D nature of the correlators at the decoupled-chain fixed point.  We proceed by  integrating out arbitrary  time differences, $-\infty < \tau_{12},\tau_{23}<\infty$,  while keeping a short-distance cutoff. We obtain  
\bea
I^{(3)}&=&\frac{3}{4\pi}\int d^3\mathbf R_1\, \langle\varepsilon_{\mathrm h}\varepsilon_{\mathrm v}(\mathbf R) \rangle_0\int_{a}^{D_y} \frac{dy_{12}}{|y_{12}|} \int_a^{D_x} \frac{dx_{23}}{|x_{23}|}\nonumber \\
&=& \frac3{\pi} \ln\left(\frac {D_x}{a}\right)\ln \left(\frac{D_y}{a}\right) I^{(1)}\;,\label{I3logs}
\eea
where we imposed both the spatial UV cutoff $a$ and IR cutoffs $D_x$ and $D_y$. Note the peculiar double logarithmic dependence, which diverges for $D_x\to \infty$ or $D_y\to\infty$. The scales $D_x$ and $D_y$ can be interpreted as being of the order of the linear system size in the $x$ and $y$ directions, respectively. As discussed in the context of  UV-IR mixing in Bose-metal-like models \cite{GorantlaPRB2021,LakePRB2022}, the result may depend on how we handle the thermodynamic limit along with the continuum limit.

We can now perform the renormalization steps on the term in Eq. (\ref{I3logs}). Rescaling $a \to a(1+dl)$, we obtain $I^{(3)} \to I^{(3)} + \delta I^{(3)} + \mathcal O(dl^2)$, where
\be
\delta I^{(3)} = -6 \ln\left(\frac {D_xD_y}{a^2}\right)  dl\; I^{(1)}\;.
\ee
The corresponding change in the perturbative expansion defined in Eq. (\ref{eq:partitionfunc}) can be written as
\bea
\frac{Z}{Z_0}&=& 1- \frac1{2\pi}\left[ g -  \ln\left(\frac {D_xD_y}{a^2}\right) g^3 dl\right] I^{(1)} \nonumber\\
&&+ \frac{g^2}{2!}I^{(2)} - \frac{g^3}{3!}I^{(3)} + \cdots \;,
\eea
where we neglected irrelevant terms stemming from $I^{(2)}$. Imposing invariance under the RG transformation, we see that the coupling is renormalized as $g \to g + dg$, which leads to  the beta function in Eq. (\ref{betafunc}).

\section{Mean-field equations \label{app:MFT}}

Here we discuss the diagonalization of the mean-field Hamiltonian and the derivation of the self-consistency equations. 

We introduce momentum modes as
\be
\gamma^b_{\mb R,\alpha} =  \sqrt{\frac{2}{N_c}}\sum_{\mathbf k \in \text{BZ}}e^{i  \mathbf k\cdot \mb R}\gamma^b_{\mathbf k, \alpha}\;,
\ee
where $\mathrm{BZ}\equiv[-\pi/d_\parallel, \pi/d_\parallel]^2$, with $d_\parallel=2a$,  stands for the Brillouin zone  of the square lattice with $N_c=L_xL_y$ unit cells. For Majorana fermions, we have the relation  $\gamma^b_{-\mathbf k, \alpha} = (\gamma^b_{\mathbf k, \alpha})^\dagger$, which allows us to restrict to modes in one half of the Brillouin zone, say $\mathrm{HBZ }\equiv [-\pi/d_\parallel, \pi/d_\parallel] \times [0, \pi/d_\parallel]$. These complex fermion operators  satisfy  $\{\gamma^{b}_{\mathbf k, \alpha}, (\gamma^{b'}_{\mathbf k^\prime, \alpha^\prime})^\dagger\} = \delta^{bb'} \delta_{\mathbf k, \mathbf k^\prime}\delta_{\alpha ,\alpha'}$.

\begin{figure}[t]
    \centering
    \includegraphics[width=\columnwidth]{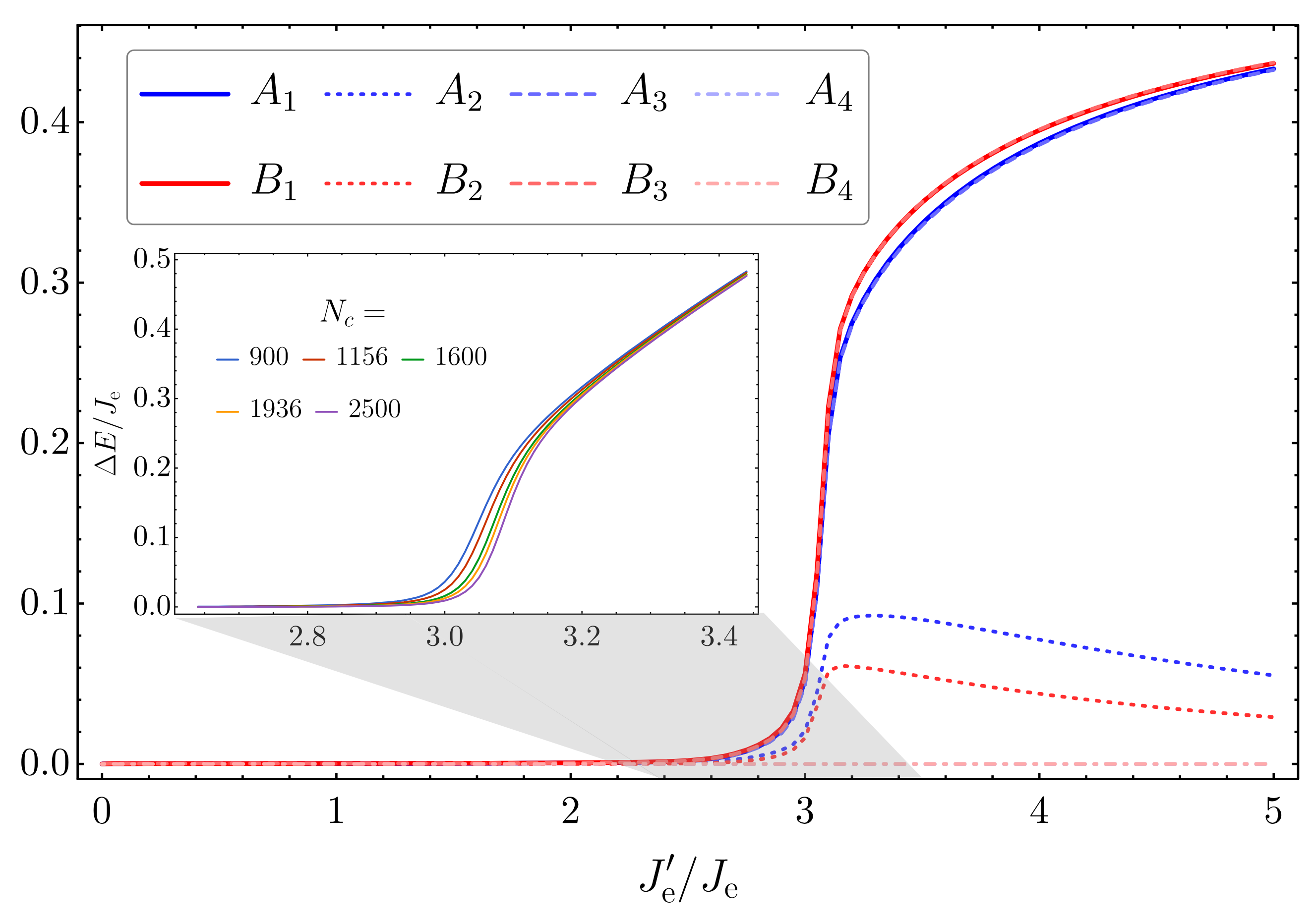}
    \caption{Results from the mean-field theory. Solutions of the eight mean-field parameters in Eqs. (\ref{eq:mfteq1}) and  (\ref{eq:mfteq2}), computed for $N_c=50 \times 50 = 2500$ unit cells. Note that the $C_4$-related parameters acquire different values for $J^\prime_\mathrm{e}/J_\mathrm{e} \gtrsim 3$. Inset: Finite-size scaling of the energy gap around $J^\prime_\mathrm e/J_\mathrm e \simeq 3$, evaluated from $N_c = 900$ to  $2500$ unit cells.}
    \label{fig:mftfig}
\end{figure}

The mean-field Hamiltonian can be cast in the form
\be
\tilde H_\mathrm{MF}  = \sum_{\mathbf k \in \mathrm{HBZ}}\Psi^\dagger_\mathbf k \mathcal H_\mathrm{MF}(\mathbf k) \Psi^{\phantom\dagger}_\mathbf k\;,
\ee
where $\Psi_\mathbf k = (\gamma^0_{\mb k,1},\; \gamma^0_{\mb k,2},\; \cdots,\gamma^1_{\mb k,3},\;\gamma^1_{\mb k,4})^T$ is an eight-component spinor and $\mathcal H_\mathrm{MF}(\mathbf k)$ is an $8\times 8$ Hermitean matrix that depends on $J_{\rm e}$, $h_{\rm e}$, $J'_{\rm e}$, as well as on the mean-field parameters defined  in Eq. (\ref{eq:mftparams}). These  parameters  are to be found by self consistency. The amplitudes of interest have the form
\be
\langle i \gamma^b_{\mb R, \alpha}\gamma^{b'}_{\mb R, \alpha+1}\rangle = \frac{1}{N_c}\sum_{\mathbf k \in \mathrm{HBZ}}\left[\langle i \gamma^b_{\mathbf k, \alpha} (\gamma^{b'}_{\mathbf k, \alpha+1})^\dagger\rangle+ \mathrm{c.c.}\right]\;.\label{selfcons}
\ee
We can find a  unitary transformation $U(\mathbf k)$ such that $U^\dagger(\mathbf k)\mathcal H_\mathrm{MF}(\mathbf k) U(\mathbf k) = \mathrm{diag}(\varepsilon_{\mathbf k, 1}, \; \varepsilon_{\mathbf k, 2}, \cdots,\;\varepsilon_{\mathbf k, 8})$. Then, the   eigenspinors $\tilde \Psi_\mathbf k \equiv (\tilde \gamma_{\mathbf k, 1},\; \tilde \gamma_{\mathbf k, 2},\;\cdots \tilde \gamma_{\mathbf k, 8})^T$  are such that $\Psi_\mathbf k = U(\mathbf k ) \tilde \Psi_\mathbf k$. The mean-field ground state is constructed by occupying all single-particle states with negative energy.  Using the band filling condition   $\langle \tilde \gamma_{\mathbf k, A}\left(
\tilde\gamma_{\mathbf k, B}\right)^\dagger \rangle =\delta_{AB} \Theta(\varepsilon_{\mathbf k, A}) \equiv  T_{AB}(\mathbf k)$ for $A,B\in\{1,\dots,8\}$, where $\Theta(x)$ is the Heaviside step function, we can rewrite Eq. (\ref{selfcons}) in a compact form:
\be
\langle i \gamma^b_{\mb R, \alpha}\gamma^{b'}_{\mb R,\alpha+ 1}\rangle = -\frac{2}{N_c}\sum_{\mathbf k \in \mathrm{HBZ}}\mathrm{tr}\left[T(\mathbf k )U^\dagger(\mathbf k)P^{b'b}_\alpha U(\mathbf k)\right]\;,
\ee
where we define the projector $P^{b'b}_\alpha$, with components  $[P^{b'b}_\alpha]_{AB} = \delta_{A,\alpha +4(b'-1)}\delta_{B, \alpha + 4(b-1)}$. Thus, the eight mean-field parameters must satisfy the following set of equations:
\begin{align}
    A_\alpha & = -\frac{2}{N_c}\sum_{\mathbf k \in \mathrm{HBZ}}\mathrm{tr}\left[T(\mathbf k )U^\dagger(\mathbf k)P^{10}_\alpha U(\mathbf k)\right]\;, \label{eq:mfteq1}\\
    B_\alpha & = -\frac{2}{N_c}\sum_{\mathbf k \in \mathrm{HBZ}}\mathrm{tr}\left[T(\mathbf k )U^\dagger(\mathbf k)P^{01}_\alpha U(\mathbf k)\right]\;.\label{eq:mfteq2}
\end{align}
Note that $U(\mb k)$ depends on $A_\alpha$ and $B_\alpha$. We solved these equations  numerically by standard iteration them until convergence is reached. 

For $J_\mathrm e \neq h_\mathrm e$, we find that the gap in both fCDW and disordered phases is stable under turning on the coupling $J_{\rm e}'$, as expected. The result for the mean-field parameters at criticality,   $J^\prime_\mathrm e = J_\mathrm e$, is shown in Fig. \ref{fig:mftfig}.  We see that $A_\alpha$ and $B_\alpha$ become nonzero only for a fairly strong interchain coupling,  $J^\prime_\mathrm e/J_\mathrm e \gtrsim 3$. The solution with nonzero hybridization breaks $C_4$ symmetry and the resulting fermionic spectrum is gapped. We have checked that finite-size effects are significant only near the critical coupling; see the inset in Fig. \ref{fig:mftfig} for the behavior of the energy gap.

\bibliographystyle{apsrev4-1_control}
\bibliography{ref}

\end{document}